\documentclass[aps,prl,twocolumn,superscriptaddress]{revtex4}
\usepackage{graphicx}
\usepackage{dcolumn}
\usepackage{bm}
\usepackage{color}
\usepackage{hyperref}
\usepackage{amsmath}
\usepackage{amssymb}
\def \gdot {\dot{\gamma}}
\def \phie {\phi_{\rm e}}

\begin{document}


\title{Shear thickening in non-Brownian suspensions: an excluded
  volume effect}

\author{Francesco Picano}
\email[]{picano@mech.kth.se}
\affiliation{Linn{\'e} FLOW Centre, KTH Mechanics, Stockholm, Sweden}
\author{Wim-Paul Breugem}
\affiliation{Laboratory for Aero \& Hydrodynamics, TU-Delft, Delft, The Netherlands}
%
%
\author{Dhrubaditya Mitra}
\affiliation{NORDITA, Royal Institute of Technology and Stockholm University,
Roslagstullsbacken 23, SE-10691 Stockholm, Sweden}
\email[]{dhruba.mitra@gmail.com}
\author{Luca Brandt}
\affiliation{Linn{\'e} FLOW Centre, KTH Mechanics, Stockholm, Sweden}
%

\date{\today}

\begin{abstract}
Shear-thickening appears as an increase of the viscosity of a dense suspension with the shear rate, sometimes sudden and violent at high volume fraction. 
Its origin for non-colloidal suspension with non negligible inertial effects is still debated.  
Here we consider a simple shear flow and demonstrate that fluid inertia   
causes a strong microstructure anisotropy that results in the formation of a \emph{shadow} region with no relative flux of particles. 
We show that shear-thickening at finite inertia can be explained as an increase of the effective volume fraction when considering the dynamically 
excluded volume due to these shadow regions.

\end{abstract}




\maketitle

The field of complex fluids is diverse and rapidly developing with potential for 
numerous relevant applications. 
Among complex fluids, on one hand we have colloidal suspensions where Brownian 
effects play an important role while inertial effects are  negligible,  
see e.g.~\cite{hoffman72,bar_jr89,marwag_jchp02,wagbra_pt09,cheetal_sci11}. 
On the other hand we have suspensions made out of larger particles, (particle radius $a > 10 \mu m$) 
where Brownian effects are negligible while inertia plays an important role.  
To be specific we shall call this second class of suspensions as non-Brownian suspensions or inertial suspensions.
Their rheology is the topic of this letter. 

Understanding the rheological properties of non-Brownian suspensions
 is not only a challenge from a  theoretical point of view~\cite{mewwag_jnnfm09,cheetal_sci11,koosetal} but has also a   
significant impact in many industrial applications, e.g.\ oil processing, cement or coal slurries~\cite{kelkel_jr91,looetal_ogst04}. 

In one of the earliest work in this field, Einstein showed that for a dilute suspension of rigid particles in a Newtonian fluid with negligible inertia 
the relative increase in effective viscosity is $\sim (5/2) \phi$ where $\phi$ is the volume fraction occupied by the particles~\cite[see e.g.,][chapter 4.11]{Bat53}. 
For higher concentrations the problem is still not well understood. 
Non-Brownian suspensions may show shear-thickening, i.e.\ an increase of effective viscosity with the 
shear-rate~\cite{stipow_arfm05,fal+lem+ber+bon+ova10}. If the volume fraction is high enough, yet below the geometrical maximum
packing, $\phi_m=0.58-0.63$, the increase of viscosity with shear-rate can be abrupt~\cite{brojae_prl09}, the so-called \emph{discontinuous shear-thickening}.

\begin{figure}[t]
\includegraphics[width=.9\linewidth]{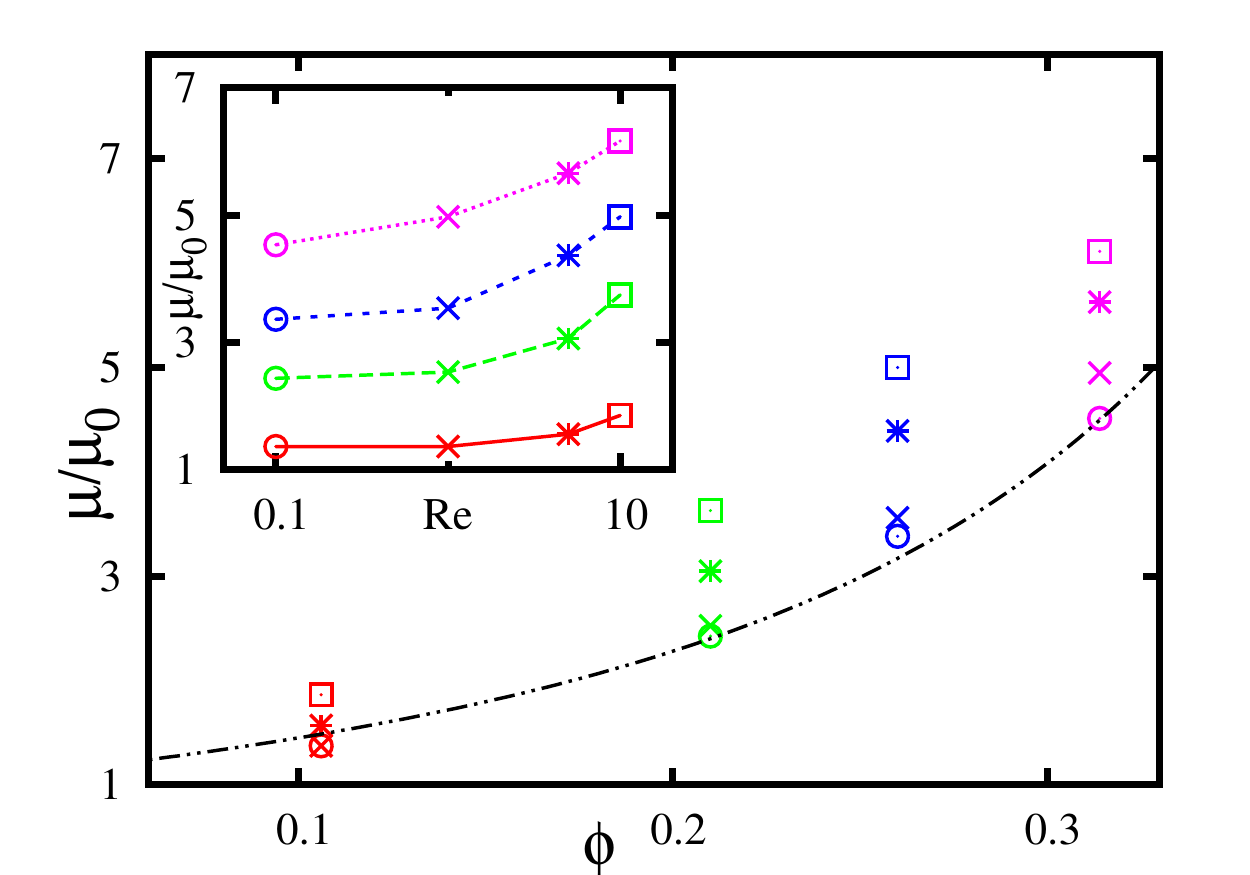}
\caption{\label{fig:1} 
\textcolor{black}{
Normalized effective viscosity $\mu/\mu_0$ vs $\phi$ for four particle Reynolds numbers $Re$;
symbols: $\circ$ $Re=0.1$,   $\times$ $Re=1$, $\ast$ $Re=5$ and $\Box$ $Re=10$;
 dash-dotted line, Eilers fit~\eqref{eq:Eilers}   with $\phi_m=0.6$ and $B=1.7$.
Inset, $\mu/\mu_0$ vs $Re$: red solid line $\phi=0.11$; long-dashed green line $\phi=0.21$; 
dashed green line $\phi=0.26$; dotted magenta line $\phi=0.315$.
}
}
\end{figure}

In this letter, we report three dimensional Direct Numerical Simulations (DNS) of a plane-Couette flow of neutrally-buoyant rigid spheres in a fluid. 
The rheology is governed by two parameters: the volume fraction, $\phi$, and the shear rate $\gdot$.
Following Ref.~\cite{stipow_arfm05}, we use a non-dimensional form of the shear rate given by the particles Reynolds number, 
$\mbox{Re}\equiv\rho \gdot a^2/\mu_0$, where $\mu_{\rm 0}$, $\rho$ are the fluid viscosity and density and $a$ is the particle
radius. 
The effective viscosity is thus function of $\phi$ and $\mbox{Re}$, $\mu=\mu_{\rm 0}\,f(\phi,\mbox{Re})$.
For the configurations investigated here, the effective viscosity, reported in Fig~\ref{fig:1},  increases as the relative strength of the inertial effects (measured by 
$\mbox{Re}$) increases; a phenomenon we call {\it inertial shear-thickening}.

\begin{figure}[tb]
\includegraphics[width=.45\columnwidth]{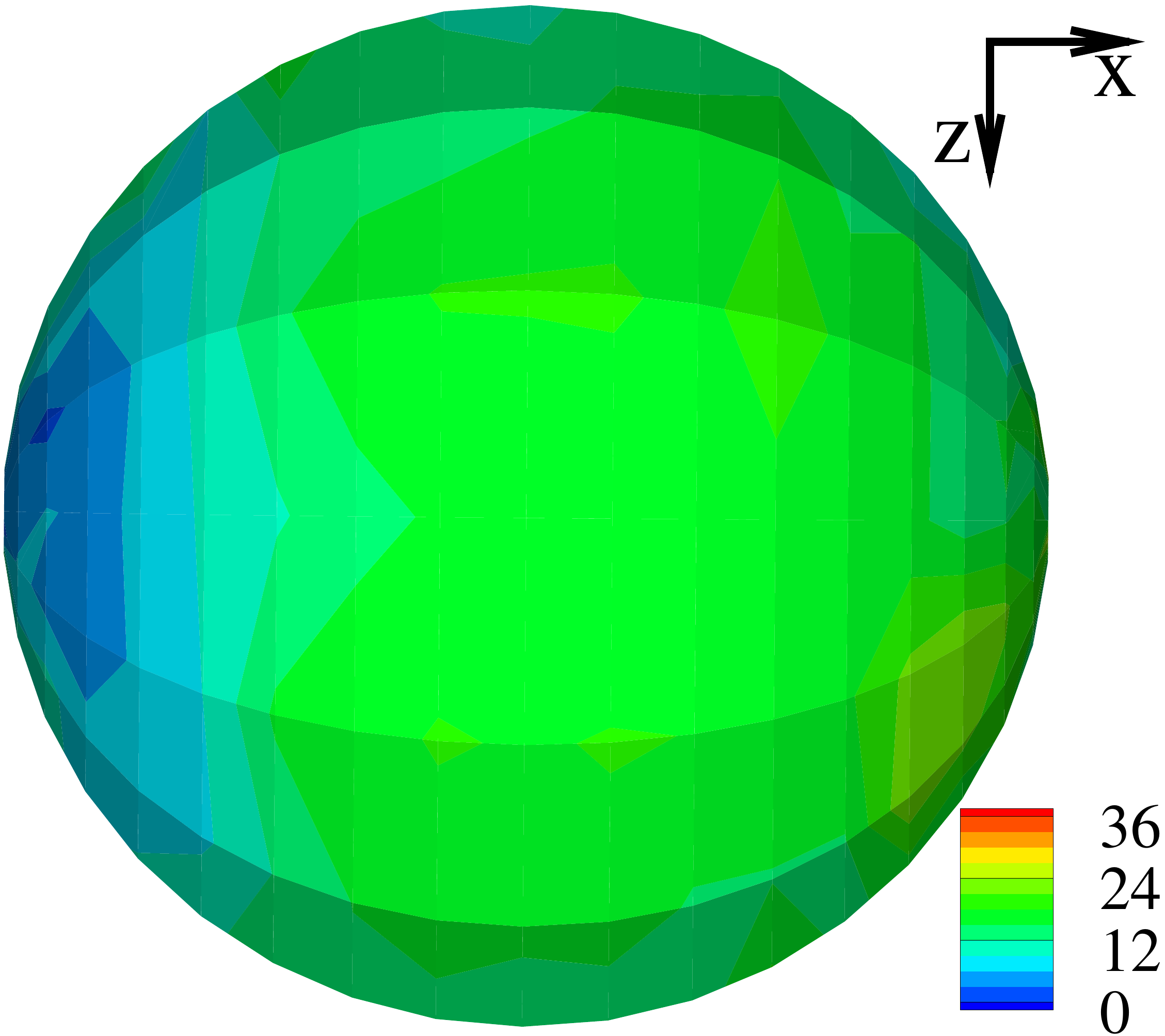}
\put(-99,99) {\Large{(a)}} 
\includegraphics[width=.45\columnwidth]{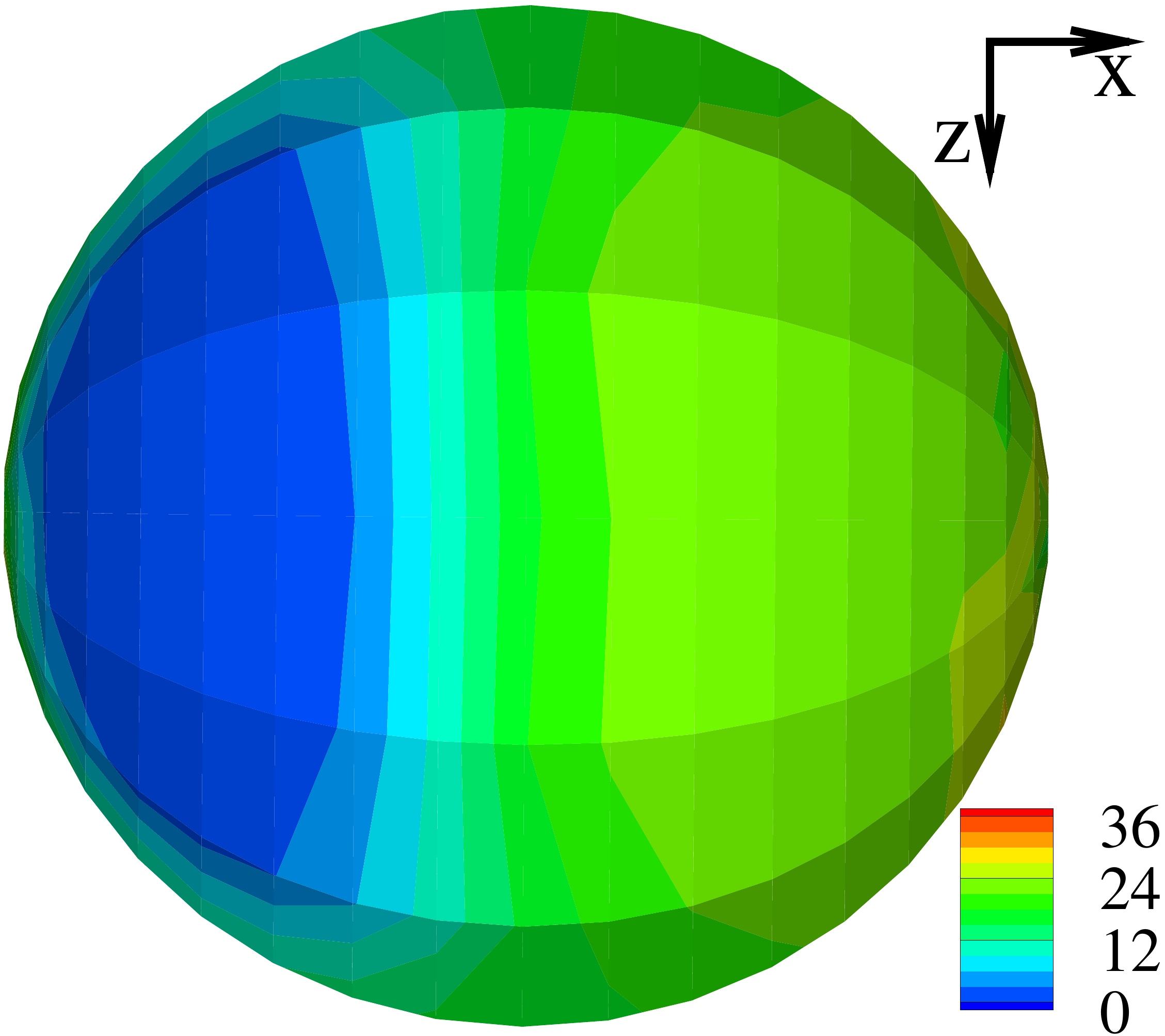}
\put(-99,99){\Large{(b)}}
\caption{\label{fig:2a}
Projection of the normalized angle-dependent pair-distribution function, $g(r\simeq2a,{\bf \hat r})$, on the wall-parallel plane (with mean flow from right to left) $x-z$ 
plane for  $Re=0.1$ (a), and  $Re=10$ (b) for $\phi=0.315$. 
}
\end{figure}

\begin{figure}[b]
\includegraphics[width=.75\columnwidth]{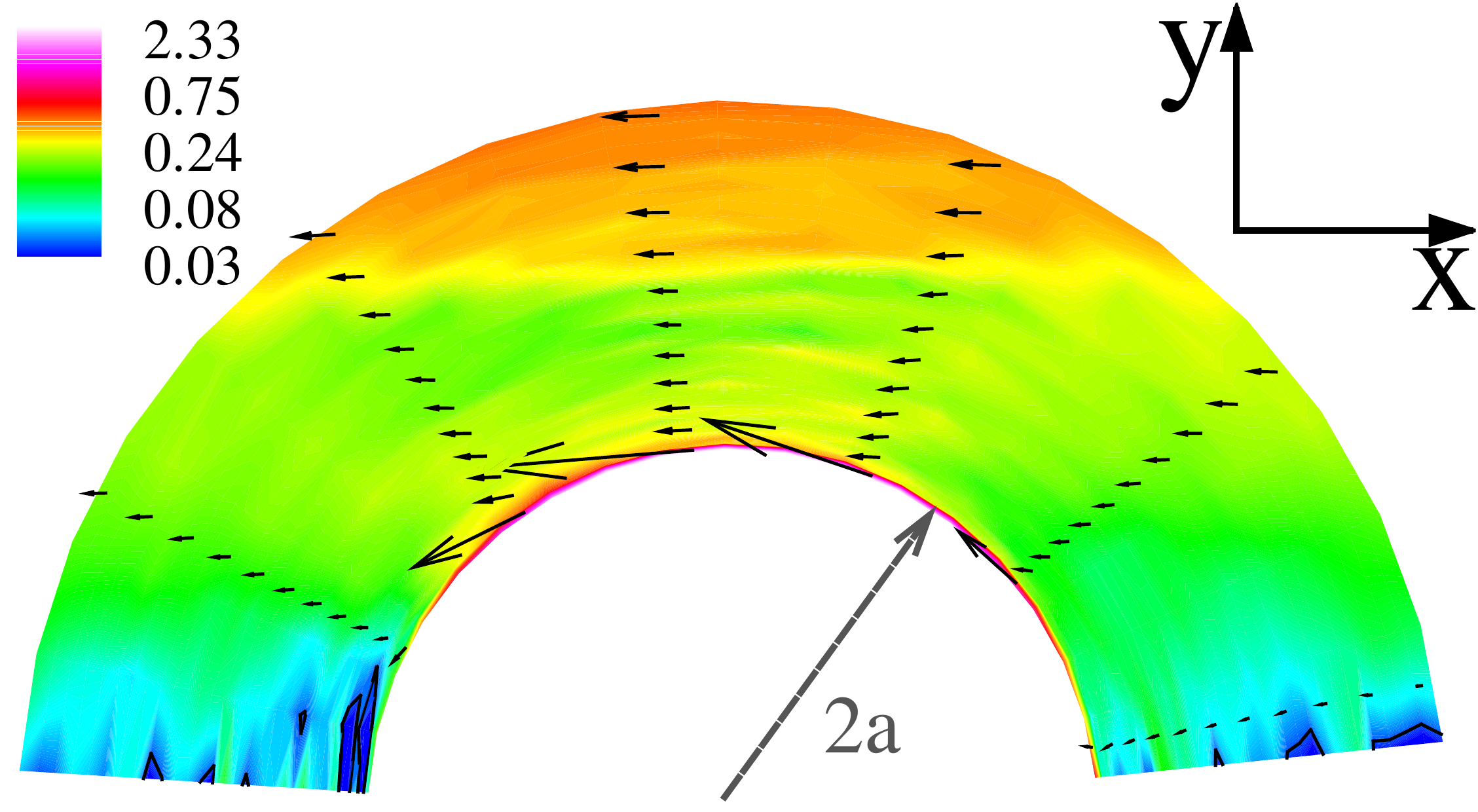} 
\put(-199,50) {\Large{(a)}} \\
\includegraphics[width=.75\columnwidth]{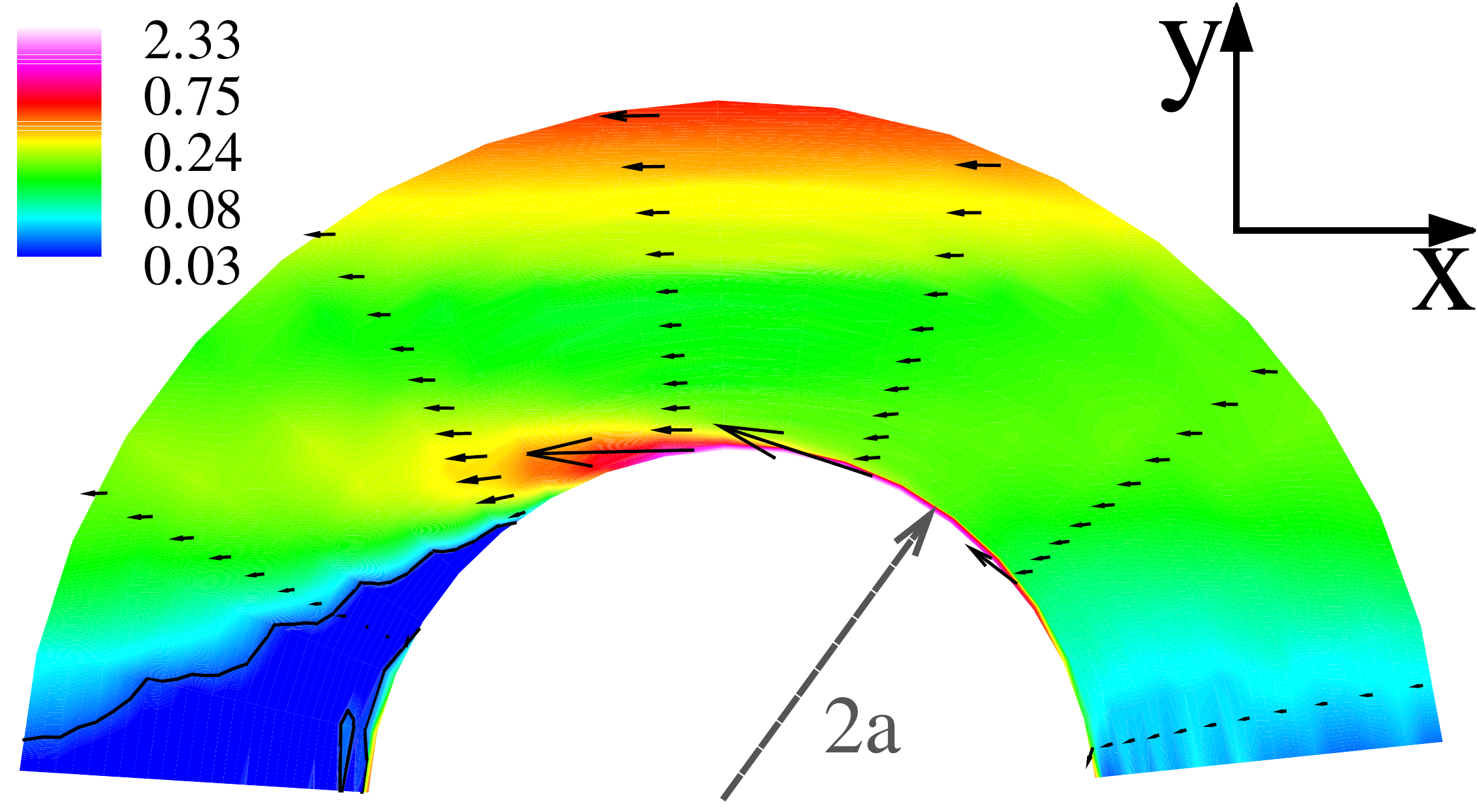}
\put(-199,50) {\Large{(b)}} 
\caption{\label{fig:flux}
Contour plot of  particle pair relative flux $\mid {\bf q} \mid$, Eq.~\eqref{eq:flux} in the shear plane $x-y$
for $\phi=0.315$: $\mbox{Re}=0.1$ (a), $\mbox{Re}=10.$ (b). Local mean flow
is from {right to left} in the horizontal direction.  The direction of ${\bf q}$ in the plane is shown by arrows.
The black contour corresponds to $\mid {\bf q} \mid = q_{\rm th}$. 
}
\end{figure}

\begin{figure}
\includegraphics[width=.85\linewidth]{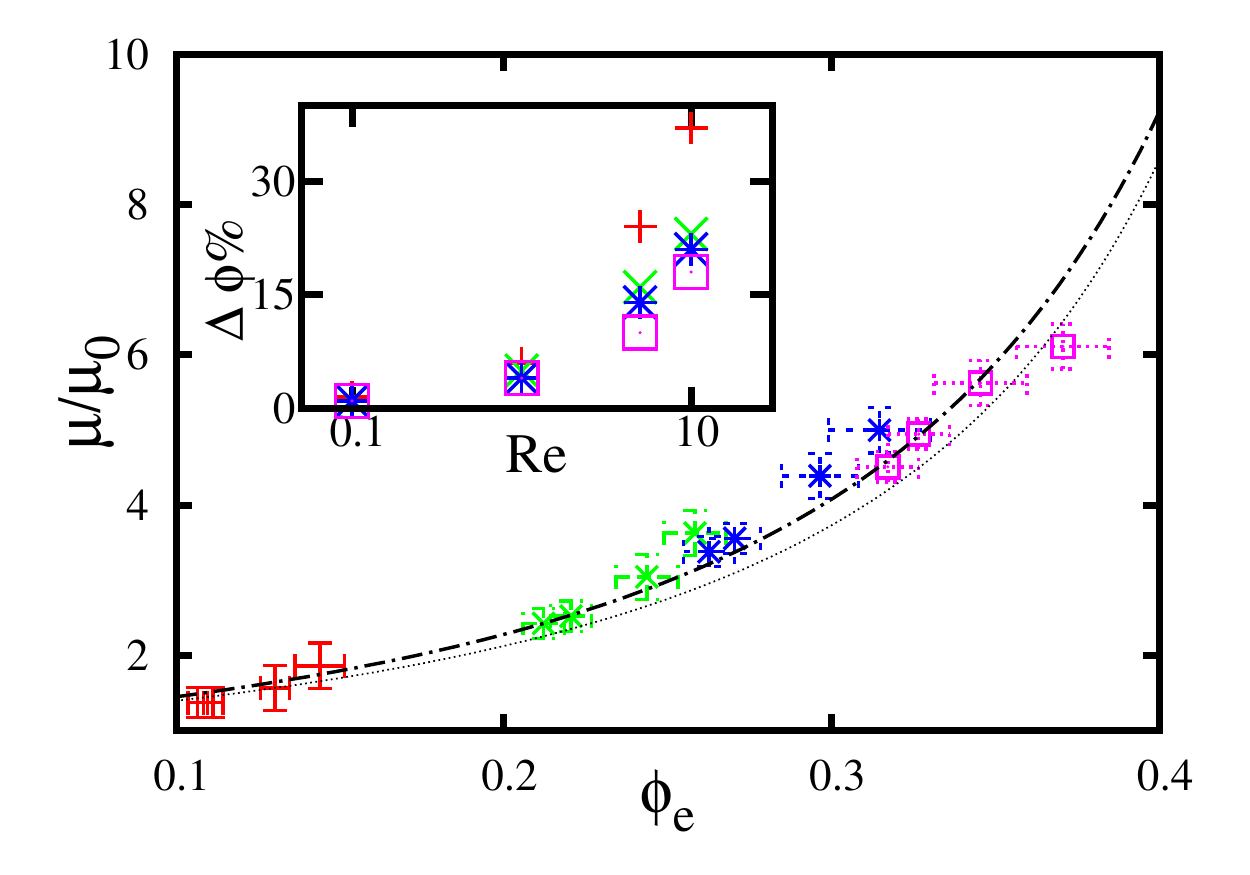}
\caption{\label{fig:3a}
Effective viscosity vs effective volume fraction $\phi_e$: red $+$ $\phi=0.11$, green $\times$ $\phi=0.21$, blue $*$ $\phi=0.26$ and magenta $\Box$ $\phi=0.315$.
Lines Eilers fit~\eqref{eq:Eilers}: dash-dotted, best fit of present data  $\phi_m=0.6$ and $B=1.7$; dotted,
fitting parameters in~\cite{zarraga_etal_jor00,sinnot_jfm03,kulmor_pof08} $\phi_m=0.58$ and $B=1.5$. 
Inset:  relative increment of the volume fraction as a function of $\mbox{Re}$.}
\end{figure}

{The relative motion of a particle pair with finite inertia in a shear flow has been studied in~\cite{kulmor_jfm08}.
These authors found that at finite Reynolds number 
the incoming particle tend to leave the reference one with a positive shift in the shear 
direction. 
Hence, we expect 
this asymmetry to affect the suspension rheology at finite $\mbox{Re}$. 
Indeed,  we find that behind a particle there exists a region with vanishing 
relative particle flux that we call \emph{shadow} region. }
We obtain an estimate of the average volume of  the shadow region in the suspension by calculating 
the pair-distribution-function, Fig~\ref{fig:2a}, and the relative flux of a pair of spheres, Fig~\ref{fig:flux}. 
We interpret the volume occupied by the shadow as an increase of the effective volume fraction; this allows us
to collapse the data for $\mu/\mu_{\rm 0}$ pertaining {four }different values of $\phi$  into one single function of the effective volume $\phi_{\rm e}$, Fig~\ref{fig:3a}. This function is well approximated by the well-known 
Eilers fit~\cite{stipow_arfm05} that is an empirical formula describing the variation of the viscosity of a suspension 
  with the volume fraction for vanishing inertia,
\begin{equation}
\frac{\mu}{\mu_{\rm 0}}=\left[1+ B \frac{\phie}{1-\phie/\phi_m} \right]^2,
\label{eq:Eilers}
\end{equation}
with $B=1.25-1.5$ and $\phi_m=0.58-0.63$ the maximum packing fraction~\cite{zarraga_etal_jor00,sinnot_jfm03,kulmor_pof08}. 
A similar collapse have been recently obtained  in granular systems under different conditions with
experimental~\cite{boetal_prl11} and numerical~\cite{tru+and+cla12} data. 
We go beyond these studies addressing the problem from a microscopical point of view and showing that this increase of the effective volume fraction is due to 
formation of {\it anisotropic} micro-structures
characterized by a angle-dependent pair-distribution function and mean relative particle flux.
Note that the existence of such
micro-structures cannot be inferred from 
isotropic, angle-averaged, observables. 
Recent investigations~\cite{bramor_jfm97,marwag_jchp02,wagbra_pt09,cheetal_sci11} have stressed the important role played by 
``hydroclusters'' in shear-thickening in Brownian (colloidal) suspensions. 
Here we elucidate the role of the particle clusters and microstructure in shear-thickening of non-Brownian suspensions with finite inertia. 
%

We numerically simulate a suspension of rigid spheres suspended in a fluid phase described by the incompressible Navier--Stokes equation.
These are solved on a Cartesian mesh in a rectangular box of size $16\,a \times 16\, a \times 10\, a$ along {the streamwise, wall-normal and spanwise directions} $(x,y,z)$, with 8 grid points per particle radius $a$. The fluid is sheared in the $x-y$ plane by imposing a constant streamwise velocity of opposite sign $U_{\rm 0} = \gdot H$, ($H=10\, a$) at the two horizontal walls ($y=\pm H/2$).
Periodic boundary conditions are imposed on the other two directions. 
A Lagrangian algorithm is used to solve for the linear and angular momentum of the spheres. We impose no slip boundary condition on the fluid
at the particle surface using the Immersed Boundary Method (IBM). Lubrication and collision models are employed to capture the interaction
between spheres when the distance between the surface of neighboring particles become smaller than the mesh size. 
The surface of each sphere is discretized by about 800 Lagrangian grid points. The code was fully validated against several classic test cases, see \cite{breugem_jcp} for more details. { Four different values of the volume fraction $\phi=0.11$, $\phi=0.21$, $\phi=0.26$ and $\phi=0.315$, and four particle 
Reynolds numbers in the range $0.1$ to $10$ are simulated.
Initially, the particles are placed at random positions, with no overlap and velocity
equal to the local fluid velocity, the laminar Couette profile.  
Statistics are collected from time $T_{\rm tr}= 20{\dot \gamma}^{-1}$
when all the simulations have reached a statistically stationary state. 
Earlier studies~\cite{pin_etal_nat05} have shown that Stokesian suspensions, although athermal, have a chaotic behaviour, hence we expect the
statistically stationary state to be independent of the choice of the initial position of the 
particles, or of the initial velocity profile. We have checked this in few representative cases.}

In Fig.~\ref{fig:1} we display the effective viscosity of the suspension, $\mu$, measured as the ratio between the tangential stress at the 
\textcolor{black}{
walls and the shear rate $\gdot$, as a function $\phi$ 
and as a function of $\mbox{Re}$ in the inset from all simulations performed.}
The effective viscosity increases with the shear rate (shear-thickens) at fixed volume fraction;  also it increases with 
the volume fraction $\phi$ at fixed $\mbox{Re}$. 
{Our results are consistent with recent numerical data in Ref.~\cite{kulmor_pof08}.}

Next we show that shear-thickening can be interpreted as an excluded volume effect. 
We first calculate the  pair distribution function $g(r,{\bf \hat r})$ that is the probability 
to find a particle pair at given distance $r$ and direction ${\bf \hat r}$  normalized by the value
for a random arrangement~\cite{mor_ra09,wagbra_pt09,cheetal_sci11}. 
%
%

In Fig.~\ref{fig:2a} we display
 $g(r,{\bf \hat r})$ at contact, $r=2 a$, in the wall-parallel $x-z$ plane (relative motion from right to left), for two different values of $\mbox{Re}$ at $\phi=0.315$ (similar behavior is observed at lower concentrations). 
The contours show that $g(r,{\bf \hat r})$ is not isotropic and the anistropy increases as the inertial effects, measured by $\mbox{Re}$, become more important. 
In particular  there exists a  small region behind the particle where there is a lower probability to find a second particle. 
Increasing the Reynolds number, the anisotropy increases.
{Though the anisotropy of  $g(r,{\bf \hat r})$ at contact has been already observed~\cite{kulmor_pof08}, its role for shear-thickening at finite $\mbox{Re}$
was not identified.
This anisotropy causes  \emph{shadow} regions with vanishing probability to find another particle in relative motion. This shadow 
acts as an increase of the effective volume fraction:  this is the geometrical volume occupied by the particles plus the volume of the shadows (the shadow is actually a property of a pair of spheres).} 

%

%
%
%
We try to estimate the volume of the shadow region by calculating the relative particle flux (relative momentum increments), defined as 
\begin{equation}
{\bf q}(r,{\bf \hat r})=g(r,{\bf \hat r})\langle \delta {\bf v}\rangle(r,{\bf \hat r}),
\label{eq:flux}
\end{equation}
 where $\langle \cdot \rangle$  denote ensemble averaging and $\delta {\bf v}$ is the relative velocity of a pair of spheres. 
The relative particle flux in the shear plane is plotted in Figure~\ref{fig:flux}. 
Clearly, the flux is largest in the region close to the surface of the sphere  (i.e. grazing incidents) and at $z>3a$ (where the mean flow determines the flux). 
Most importantly, there exists a region behind a sphere where this flux reaches a minimum value, close to zero, for
$\mbox{Re}\ge1$.  We call this region the shadow region. 
To estimate the volume occupied by the shadow region, we select a  threshold value $q_{\rm th}=0.03$ (black contour in figure~\ref{fig:flux}) and
calculate the volume of the region where $\mid {\bf q} \mid \leq q_{\rm th}$.
This volume, function of the particle Reynolds number and volume fraction, ${\cal V}_{\rm d}(\mbox{Re},\phi)$, 
is the relative increase of the suspension excluded volume  $\Delta \phi/\phi={\cal V}_{\rm d}/{\cal V}_{\rm g}$, where ${\cal V}_g=4\pi (2\,a)^3/3$.
\textcolor{black}{The relative increment of the volume fraction is displayed as a function of ${\mbox Re}$ for the four
different values of $\phi$ in the inset of Fig.~\ref{fig:3a}. }
The increase of the volume fraction is significant, of the order of $10\%$  for $Re\ge1$. 
At fixed $\mbox{Re}$, the relative increase of the effective volume fraction 
decreases marginally at larger $\phi$ since collisions among particles are more 
frequent and deflect the particle trajectories reducing the size of the shadow region.         
The values of the effective viscosity, $\mu/\mu_0$, in the range of $\phi$ and $\mbox{Re}$ considered can be collapsed to
an universal curve using the effective volume fraction $\phi_{\rm e}(\phi,Re)=\phi + \Delta \phi (\mbox{Re})$, see Fig~\ref{fig:3a},  
where we also plot the Eilers fit~\eqref{eq:Eilers}~\cite{zarraga_etal_jor00,sinnot_jfm03,kulmor_pof08}, valid for suspensions of vanishing inertia.
We indeed find a good agreement between Eilers Fit and our data given the crude nature of the estimate of the relative increase of volume fraction~\footnote{The  choice of $q_{\rm th}$ is arbitrary, however no significant changes of the results are observed even if $q_{\rm th}$ is changed by about $50\%$.}.
\begin{figure}[t]
\includegraphics[width=.78\columnwidth]{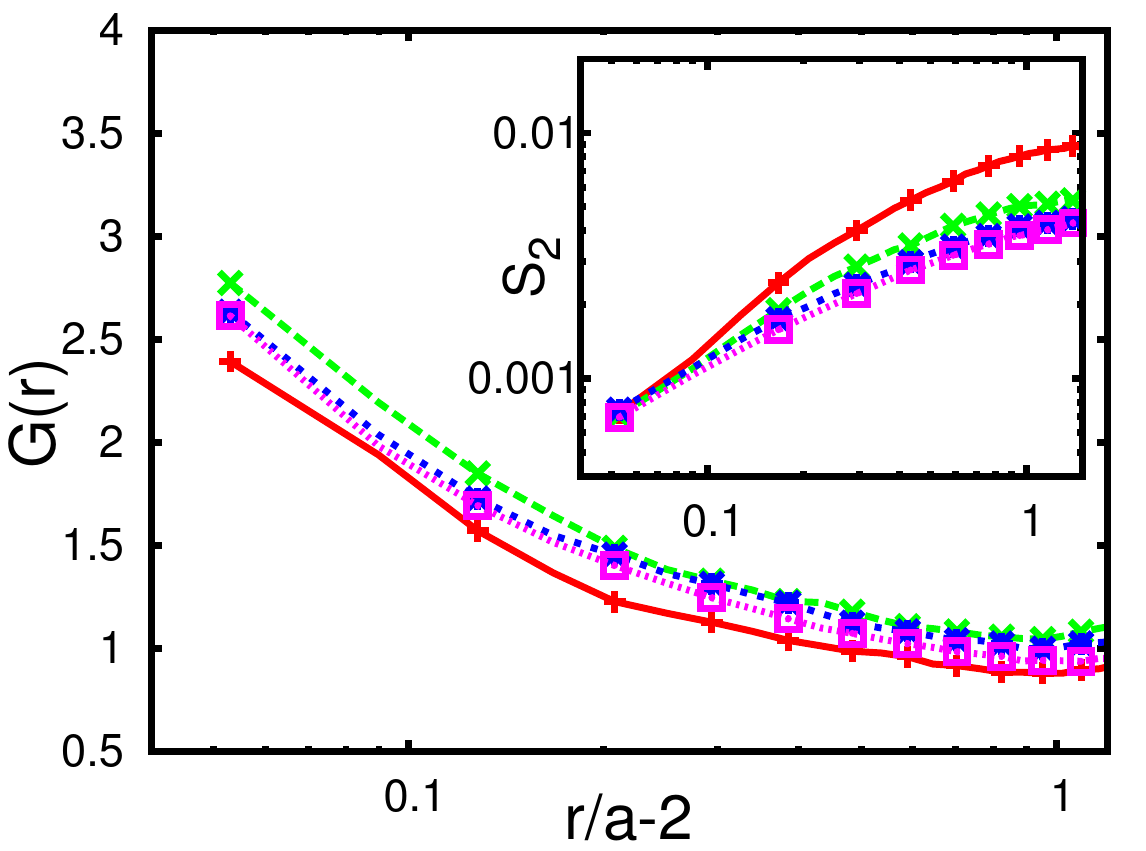} 
\put(-155,30){\Large{(a)}} \\
\includegraphics[width=.78\columnwidth]{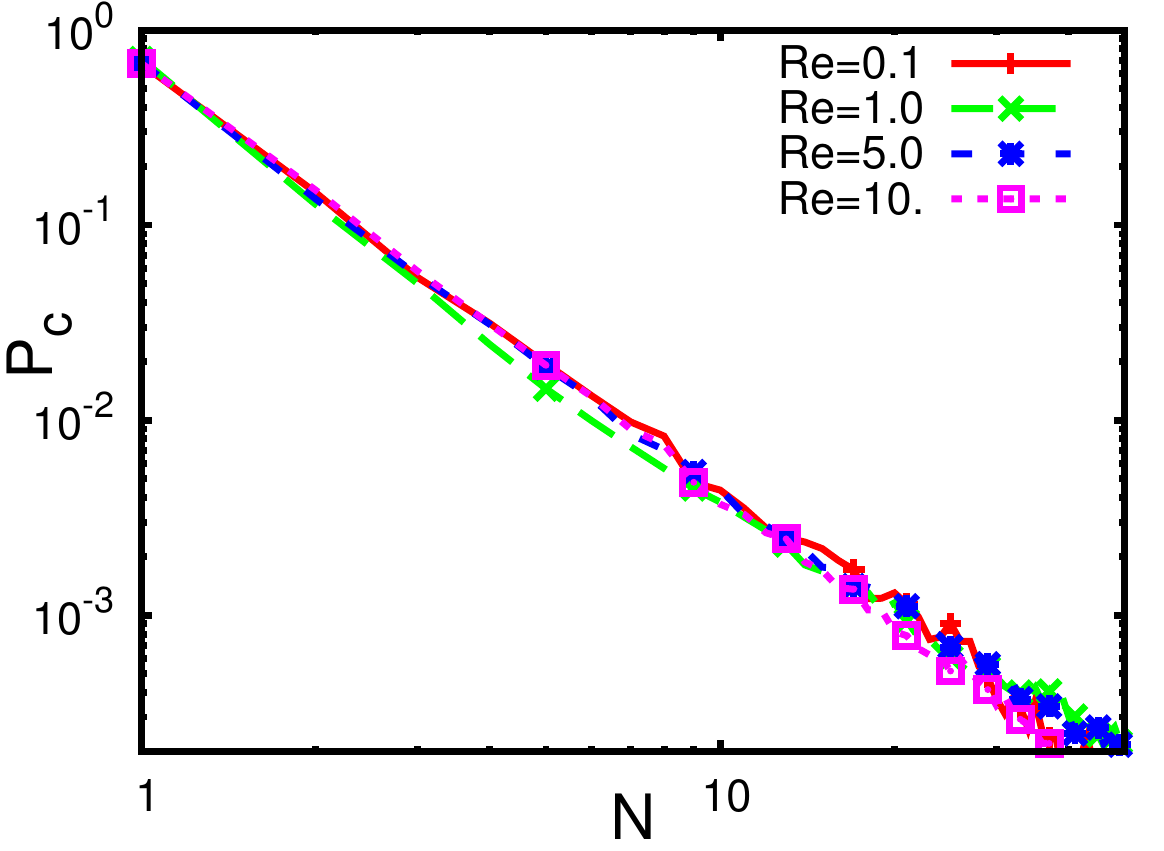}
\put(-155,30){\Large{(b)}}
\caption{\label{fig:4a}
(a) The pair-distribution function averaged over unit sphere, $G(r) \equiv (1/4\pi) \int g(r, \hat{r}) d\Omega $ vs $r/a -2$ for
$\phi=0.315$ and several values of $\mbox{Re}$: red $+$ $\mbox{Re}=0.1$, green $\times$ $\mbox{Re}=1.$, blue $\ast$ $\mbox{Re}=5.$ and
magenta $\Box$ $\mbox{Re}=10.$. Inset: Second order structure function of longitudinal velocity differences of the spheres
vs $r/a -2$. 
(b)  Probability distribution function of the number of clusters formed by $N$ spheres.}
\end{figure}
%
%
%
\begin{figure}[tb]
\includegraphics[width=.8\columnwidth]{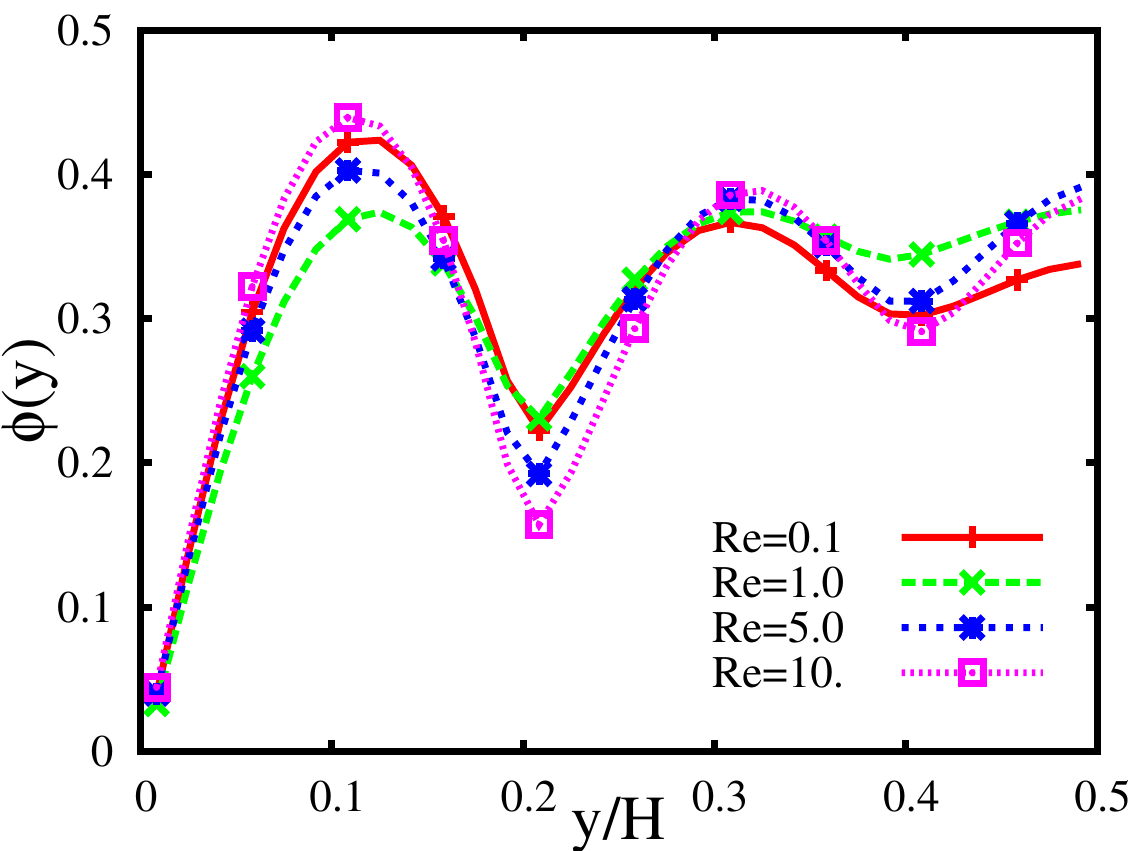}
\put(-155,30){\Large{(a)}} \\
\includegraphics[width=.8\columnwidth]{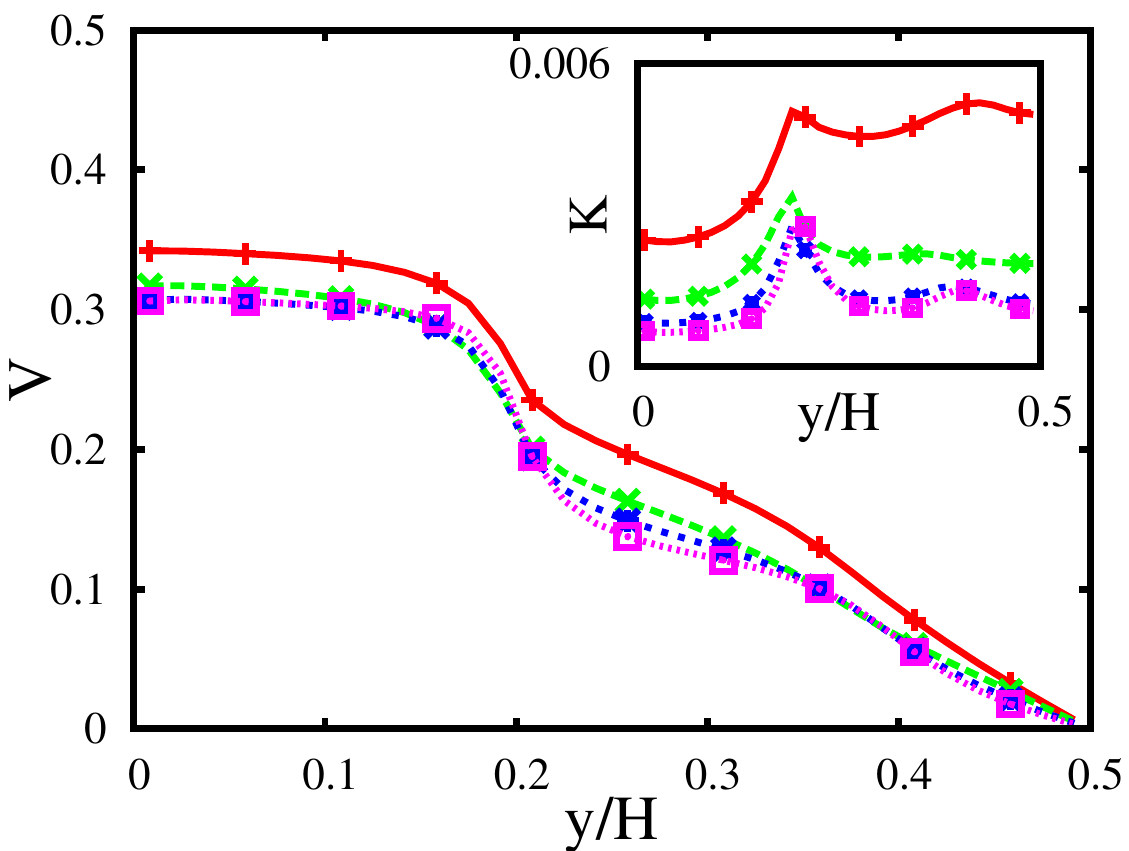}
\put(-155,30){\Large{(b)}}
\caption{\label{fig:5a}
 Wall-normal profile of (a)  the average local volume fraction, and 
 (b) average particle velocity, $V(y)$, for $\phi=0.315$. 
 The inset shows $K \equiv v^2_{\rm rms}$ versus the wall distance $y/H$. Symbols are as in Fig~\ref{fig:4a}.}
\end{figure}
%
%

We stress that the increase in effective volume fraction is essentially due to the formation of anistropic microstructures, as already seen in Figs.~\ref{fig:2a} and \ref{fig:flux}.  
We present three further evidences to support this claim:
(a)  We plot in Fig.~\ref{fig:4a}a the pair-distribution function averaged over the solid angle $G(r) \equiv (1/4\pi) \int g(r, \hat{r}) d\Omega $.
Although clustering at small distance is clearly present ($G(r) > 1$ for small $r$), no significant change is observed with $\mbox{Re}$. 
(b) We report the  second order structure function of the longitudinal particle velocity difference,
 $S_{\rm 2}(r) \equiv (1/4\pi) \int \langle \delta v_{\parallel}({\bf r})^2\rangle d\Omega$,  as an inset in figure~\ref{fig:4a}a.
  $\delta v_{\parallel} ({\bf r}) =[{\bf v}_{\rm P} - {\bf v}_{\rm Q} ]\cdot {\bf \hat r}$, 
where ${\bf v}_{\rm P}$ and ${\bf v}_{\rm Q}$ are the velocities of the P-th and Q-th particle separated by a distance ${\bf r}$.
Similar to $G(r)$,  $S_{\rm 2}(r)$ does not show any significant change at small separation $r$ when increasing $\mbox{Re}$. 
(c) We display in Fig.~\ref{fig:4a}b the probability distribution function of the number of clusters containing $N$ spheres, $P_{\rm c}(N)$.
 Particles are considered to belong to the same cluster if their gap distance is less than 2\% of $a$.  
 We find that $P_{\rm c}(N)\sim N^{-2} $, i.e., there exists finite probability to find large aggregates,
as observed for shear-thickening colloidal suspensions~\cite{wagbra_pt09,cheetal_sci11}.   However $P_{\rm c}(N)$ does not change as a function of $\mbox{Re}$.
Hence, though hydroclusters are present, 
we do not observe a direct connection between formation or growth
 of clusters and inertial shear-thickening.

{
The wall-normal profile of the local mean volume fraction $\varphi(y)$, Fig.~\ref{fig:5a}a, shows that particles tend to form layers due to the confinement from the wall, see also the mean particle velocity $V(y)$ in~\ref{fig:5a}b. Again layering 
does not show a monotonic behavior with $\mbox{Re}$.
{Consistently with~\cite{kulmor_pof08}, } single-point
particle velocity fluctuations decrease with the inertia, as shown in the inset of Fig.~\ref{fig:5a}b.
The system appears more stable, with a more ordered structure and fewer particles jumping among the layers when increasing $\mbox{Re}$.}

The decrease of the fluctuation level and the increase of the ordering is consistent with the idea of an 
increasing effective volume fraction at high shear-rates:
the system tends to freeze as there is less available space for the particle motion. 
{
We conjecture that if} the effective volume fraction approaches the critical packing, the system would \emph{jam}. 
Hence we {may hypothesize} that  the \emph{discontinuous shear-thickening} observed at high 
concentrations, higher than those simulated here, yet below the geometrical maximum packing  $\phi_m$, can be interpreted as an increase of the effective volume fraction above $\phi_m$,
$\phi<\phi_m\le\phi_e(\mbox{Re})$. 
This behavior might appear as heterogeneity in space with part of the system jammed at large shear rates $\gdot$~\cite{fal+lem+ber+bon+ova10}.
\textcolor{black}{Nonetheless, it should be remarked that the anisotropic shape of the shadow regions may also change the maximum packing fraction 
$\phi_m$, e.g.~\cite{Donev13022004}}
We hope our work will promote new 
research on the dynamics of the microstructure in these regimes.
 
To conclude, we show in this letter that inertial shear-thickening in non-Brownian suspensions can be understood in terms of an increase of the 
effective volume fraction of the suspension. 
{The presence of inertia modifies the relative particle motion (development of shadow region) increasing the level of mutual interactions (increased excluded volume). We show that this is the main effect of inertia since the effective viscosity follows a relation that
holds for the case of zero inertia, eq.~\eqref{eq:Eilers},  when considering the effective volume fraction $\phi_e$.}\\


\begin{acknowledgements}
DM thanks Pinaki Chaudhuri and John Wettlaufer for useful discussions and the Swedish Research
Council for the support through grant no.\ 2011-5423. Computer time provided by SNIC, Sweden, and CASPUR, Italy  (\emph{std12-084} grant) is gratefully acknowledged. 
\end{acknowledgements}

\end{document}